\documentclass[aps,pre,twocolumn,epsfig,epsf,showpacs]{revtex4}
\usepackage{graphicx}
\usepackage{amssymb}
\usepackage{amsmath}

\begin{document}
\widetext

\title{Low Temperature Properties of the Random Field Potts Chain}

\author{Raja Paul{$^1$}, Mikko Alava{$^2$}, Heiko Rieger{$^1$}}

\affiliation{$1$Theoretische Physik, Universit\"at des Saarlandes, 
66041 Saarbr\"ucken, Germany}

\affiliation{
${\ddagger}$Helsinki University of Techn., Lab. of Physics, 
P.O.Box 1100, 02015 HUT, Finland}


\thispagestyle{empty}

\begin{abstract}
The random field $q$-States Potts model is investigated using
exact groundstates and finite-temperature transfer matrix 
calculations. It is found that the domain structure and
the Zeeman energy of the domains resembles for general
$q$ the random field Ising case ($q=2$). This is also
the expected outcome based on a random-walk picture of the
groundstate. The domain size distribution is 
exponential, and the scaling of the average
domain size with the disorder strength is similar for

$q$ arbitrary. The zero-temperature properties are
compared to the equilibrium spin states at small
temperatures, to investigate the effect of local random
field fluctuations that imply locally degenerate regions.
The response to field perturbations ('chaos')
and the susceptibility are investigated. In particular
for the chaos exponent it is found to be 1
for $q = 2,\ldots,5$. Finally for $q=2$ (Ising case) the
domain length distribution is studied for correlated random fields.

\end{abstract}

\maketitle

\section{introduction}

Adding disorder to random magnets creates new effects one the prime
examples being the random field Ising model (RFIM)
\cite{review,rfim-rieger}.  Since the randomness couples directly to the
order parameter one has interesting physics in the groundstate (GS)
properties \cite{rfim-gs,gs1,gs2,gs3}, already.  With a finite
temperature the disordering effects of entropy and local fields compete.

The purpose of this article is to explore the generalization or
extension of the one-dimensional RFIM to the one-dimensional
random-field Potts model.  The 1d RFPM is defined on a chain of $L$
spins (with periodic boundary conditions being used here) where each
spin site can be in one of $q$ states. The Hamiltonian is given by

\begin{equation}
{\mathcal{H}} = -J\sum_{\langle i,j\rangle}(q\delta_{\vec{\sigma_i},\vec{\sigma_j}} - 1) - \sum_i h_i(q\delta_{\vec{\sigma_i},\vec{\alpha_i}} - 1)
\label{eq1}
\end{equation}

\noindent where $\langle i,j\rangle$ are nearest-neighbour pairs and
$\delta_{\sigma,{\sigma}'}$ the Kronecker-delta,
 i.e., $\delta_{\sigma,{\sigma}'}= 1$ for $\sigma={\sigma}'$ and
$\delta_{\sigma,{\sigma}'}= 0$ otherwise . The vector 
$\vec{\alpha}_i$ is a unit vector pointing randomly in one of the $q$ spin 
directions and $h_i$ is the local field strength which is chosen either 
constant or randomly distributed. In the numerical computations
presented below we have used a Gaussian distribution for the random
fields with $\langle h_i \rangle = 0$ and $\langle {h_i}^2 \rangle = {h_r}^2,$ such that $h_r/J$ is a measure for the strength of the disorder,
 except at finite $T$ where fields are uniformly distributed between 
[$-\delta,+\delta$]. In the following we denote the one-dimensional
versions of the RFIM and the RFPM with RFIC(random field Ising chain)
and RFPC(random field Potts chain).

Our approach is to explore the RFPC by three different techniques.  We
compute the exact GS by using a shortest path method\cite{gs1,gs2,gs3}.
This is a generalization of an idea that allows to find the GS of the
RFIC, and can be understood as the zero-temperature limit of the
transfer matrix computation of the equilibrium spin state.  This is also
employed here, to compare that state with the GS. 
Finally we resort to a qualitative description in terms
of a random-walk (RW) picture of the groundstate. T
his is also an extension of the RFIC case, where an
earlier paper\cite{Schroeder01} presented a RW description of the exact
groundstates.

There is rather little work on the RFPC with the exception
of trivial field distributions \cite{chains,Binder}.
Thus it is worth to recapitulate some of the features of the RFIC
or the reasons why it still receives some attention. Due to the
fact that the model is one-dimensional one could expect that
the basic magnetization properties are more or less trivial.
It turns out to be so, however, that in particular for binary
random field distributions ($h_i = \pm h_r$) these have very peculiar
properties\cite{Bruinsma83,Andelman86}. The equation for the local 
magnetization can be considered as a 1d dynamical system with, for 
suitable parameters, a multifractal probability distribution\cite{Bene88}.
For the RFPC this is perhaps also the case, generalized to higher
dimensions if $q>2$. In this article we omit such considerations for the
sake of the equilibrium and groundstate properties. 

One of the points of an exact solution for the GS is that
it can be investigated how exactly the introduction of a small,
finite temperature breaks up the GS. For the RFPC it is more
or less trivial that the {\em overlap} between the GS and the
$T>0$ state stays close to one, contrary to e.g. spin glasses
where this question may still be open to a debate. 
However, it is of interest to study how exactly the GS is modified by the
``easy'' excitations. These are, as discussed below, related
to the almost degenerate regions of the GS. The equilibrium state 
is also of interest as the asymptotic one for out-of-equilibrium
processes. In such one-dimensional systems the domain walls undergo
activated dynamics, in particular the RFIC ones perform Sinai walks
\cite{Bouchaud90,LeDoussal}.

The structure of the rest of the paper is as follows. First we
overview, in section II, the theoretical expectations for the
groundstate properties and outline qualitatively the consequences
for perturbations from a given GS, whether by temperature or by
changing the local fields (by random shifts or by an uniform
applied field). Section III is devoted to exact numerical studies
of the RFPC. In Section IV the temperature is switched on, and
the effect of entropy on the domain structure is analyzed. Section
V contains a brief numerical study of the effect of correlated fields.
Finally, Section VI finishes the paper with conclusions. The
numerical techniques are discussed in the Appendices.

\section{Random Walks: decomposition of the groundstate}

We have earlier presented a way to divide the sequence of random
fields $h_i$ for the RFIC in such a way so that one can understand
the ensuing domain structure \cite{Schroeder01}. The idea is to
look at trial random walks: start from a test site, and follow
the sum of the random fields left/right (an arbitrary choice).
Each of such trials constitutes either an absorbing or non-absorbing
one. The terms describe the fact that trials are random walks with
absorbing boundaries. One boundary results in an excursion
of the RW which is such that enough Zeeman energy ($2J$) is accumulated
to form a domain, but if the other one is met first the try is
finished and a new one is started. Since the RW's are independent,
the GS factorizes. Mathematically such excursions are given by the
following rules. An $absorbing$ excursion ($ae$) is a sequence 
$\mathcal{S}$ of 
spins starting at some lattice site $i$ and ending at $j\geq i$, with the 
field sum $|\sum_{i\in{\mathcal{S}}}h_i|$ for the first time becoming 
greater or equal to $2J$. On the other hand a sequence $\mathcal{S'}$ of 
spins from $i$ to $j\geq i$ is a \mbox{\it{non-absorbing}} excursion ($nae$)
if $0<\bar{\sigma}\sum_{l=i}^jh_l< 2J$ $\forall i< k< j$ ; 
where $\bar{\sigma} = \pm 1$ is the orientation of the spins within the 
preceding $absorbing$ excursion. 

\begin{figure}\hspace{-3ex}
\includegraphics[width=\linewidth]
{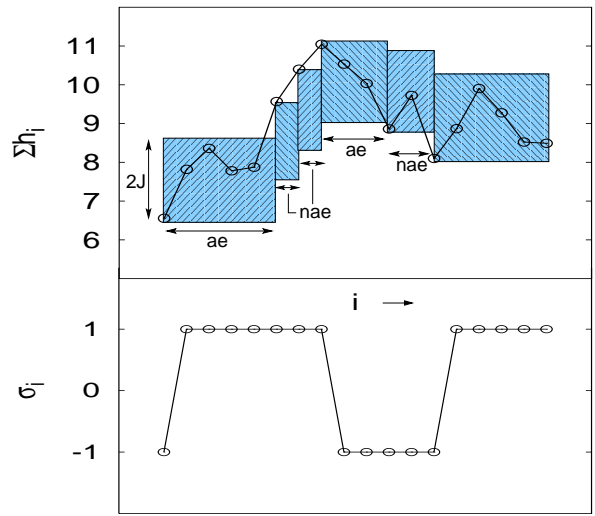}
\caption{Random walk picture for the RFIC domains. The upper part of the
 figure shows the field sum $\sum h_i$ at each of the spin sites
 (starting from the left) and the lower part depicts the corresponding
 domain structure. The random walk starts a $non-absorbing$ (nae)
 excursion as
 soon as it leaves the shaded region of height $2J$(for details
 see text).}
\label{fig0}
\end{figure}

As shown in Figure \ref{fig0} for each domain one has at least one such absorbing RW (it can
contain more than one such large fluctuation) and it will be bounded by
the next opposite absorbing RW's on each side. The rules to be followed are:
(1) Determine an absorbing excursion ${\mathcal{S}}_0$ for a given 
field configuration. If it starts at site $i_0$ and ends at $j_0$, 
and $\bar{\sigma}$ is the sign of its field sum, then $\sigma_k = 
\bar{\sigma}$  $\forall k\in {\mathcal{S}}_0$. (2) Starting from 
$j_0+1$ find all \mbox{\it{non-absorbing}} excursions until the next 
absorbing excursion ${\mathcal{S}}_1$ (from $i_1$ to $j_1$) is found, 
whose field sum is by definition opposite in sign to the preceding one. 
The sites $k$ belonging to the \mbox{\it{non-absorbing}} excursions have
the same orientation $\sigma_k = \bar{\sigma}$ as those within 
${\mathcal{S}}_0$. The orientation of the spins at sites $l$ within 
${\mathcal{S}}_1$ is opposite to the later one, $\sigma_l = -\bar{\sigma}$. 
(3) Starting at $j_1+1$ the search (2)  for the next absorbing excursion 
then leads to the overall GS.

This picture explains also, for binary random fields $h_i = \pm h$, the
degeneracy of the RFIC. For our purposes it is more important to note
that for any kind of perturbations it implies that there are three
processes, domain wall-shifts, destruction and creation of domains, that
follow naturally from the sequence of RW's. If one changes the local
random fields some of the absorbing excursions become non-absorbing, and
thus domains are destructed while new ones may ensue if some
non-absorbing ones become absorbing. These are in principle easy and in
practice cumbersome (since one needs to be concerned with the exact
first passage properties of the perturbed field distributions or random
walk step length distributions) to compute for each process. Moreover,
the normal outcome is that the domain-walls are shifted. The major
results that follow for the RFIC from the RW picture concern the domain
structure and optimization. The domain magnetic (Zeeman) energy scales
linearly with domain size in contrast to the Imry-Ma
argument\cite{Imry75}. The domain size distribution is exponential, and
for $h_r \ll 1$ the Imry-Ma result $\langle l_d \rangle \sim 1/h_r^2$ is
recovered since both the contributions from the non-absorbing walks and
the absorbing ones that contribute to the domain length scale with
$1/h_r^2$. Thus for the RFPC it will be interesting to compare both the
domain structure and the Zeeman energy,
\begin{equation}
 H_Z = \sum_{i\in {\mathcal{D}}}
h_i(q\delta_{\vec{\sigma_i},{\vec{\alpha_i}}}-1).
\label{eq2}
\end{equation}

For the Potts model one can use the same thought experiment: assume
the spin $\sigma_i=q''$, and look for ``$absorbing$ excursions''. 
Now, in contrast to the $q=2$-case the general RFPC is more complicated.
 For each of the other $q' \neq q''$ one can
follow, analogously to the RFIC, the random walk that results from
three steps: 1) null step (local field aligned to other direction
than $q'$ and $q''$), 2) positive and 3) negative steps. One has 
$q-1$ such trial walks, with the elementary step directions
being {\em correlated} or shared. These follow ``partial'' sums
of local random fields, according to the rules 1) - 3).
Thus an analogy of the RFIC results
in studying the first passage properties of a $q-1$ dimensional space.
Either one of the coordinates is increased, or all of them are decreased
($q''$-step) at the same time. One is interested in the first-passage
through one of the sides of a hyper-cube in $q-1$ -dimensions, ie. across
the plane where $q' = 2J$. The walk starts in the immediate vicinity
of the corner $(0,\dots,0$) of the hyper-cube. The simplest complication
that arises over the $q=2$, RFIC case is that for each trial walk that
compares any of the spin orientations and the original one there are 
``empty'' steps consisting of local fields aligned with 
the other $q-2$ possibilities. This implies that the typical 
domain length might simply scale with $q$, as borne out later by
the numerics. 

There are two further complications compared to the RFIC, both
resulting from the fact that the optimization can not be done
as straightforwardly as in the RFIC
due to the factorization of the landscape
to absorbing and non-absorbing excursions. Namely, for $q>2$ the
process is slightly non-local. Consider the $q=3$ case, and assume
that spin $i$ has $q'=1$. If a large fluctuation is found by
following the appropriate partial sum of the $q'=2$-orientation,
and the next one is of $q'=3$-kind, one has to consider whether
it is energetically more favorable to create a $2$-domain, a 
$3$-domain, or both, since the number of domain walls created
can be lowered by omitting in this example the $2$-domain. One
can also use the argument the other way: if a large $3$-fluctuation
exists, it is possible to create a $2$-domain since one of the
domain walls is free. The following structure
1 1 1 {\bf{1}} 3  3 3  of total energy $E_i$, is changed into
1 1 1 {\bf{2}} 3 3 3   of energy $E_f$ by flipping a single spin 
$\bf{1}\rightarrow \bf{3}$ (the local random field $h_i = h$ directed 
along $\vec{\alpha}$ = 3 at the flipped site). The latter configuration 
would be more favourable than the former one if, $E_f<E_i$, yielding $h>J$. 
This shows that the minimum Zeeman energy of a domain of length 1 should 
at least be  $J(q-1)$. 

\section{Ground State Properties}
\subsection{Correlations among the successive domains and the
evolution with increasing disorder}

As noted above the sequence of domains is complicated due to the
joint optimization. The easiest quantity is to compare the probability
that the third domain from the original one is of the same orientation
as the first one. 
We calculate the probability $P_{\sigma}(q)$  of finding the every third 
domain (${\mathcal{D}}_{i+2}$) to be the same or different as the 
first (${\mathcal{D}}_{i}$) one. In absence of any correlations
among the successive domains one should expect the probability 
${P}_{\sigma}^{nc}(q)$ of obtaining ${\mathcal{D}}_{i+2}$ same or 
different with respect to ${\mathcal{D}}_{i}$ to be,

\begin{eqnarray}
{P}_{\sigma}^{nc}(q) = 
\left\{\begin{array}{cll}
\frac{1}{q-1} &,\mathrm{ for } &{\mathcal{D}}_i = {\mathcal{D}}_{i+2}\\
\\
\frac{q-2}{q-1} &,\mathrm{ for }& {\mathcal{D}}_i \neq {\mathcal{D}}_{i+2}
\end{array}\right. 
\label{eq3}
\end{eqnarray}

\begin{figure}\hspace{-3ex}
\includegraphics[width=\linewidth]
{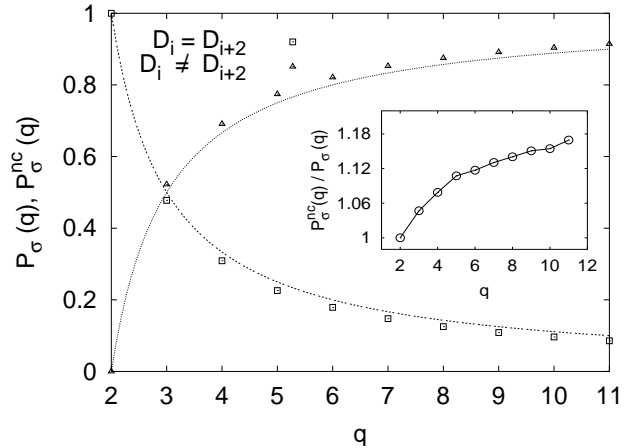}
\caption{{\bf{Left: }}Probability of obtaining every third domain 
${\mathcal{D}}_{i+2}$ same or of  different as the first ${\mathcal{D}}_{i}$.
 The continuous lines represents the no-correlation probability function 
${P}_{\sigma}^{nc}(q)$. {\bf{Inset :}} The ratio of 
${{P}_{\sigma}^{nc}(q)}/{P_{\sigma}(q)}$ means the third domain 
is the same as the first one. 
The system size $L = 10^5$ and the number of configurations 20000.}
\label{fig1}
\end{figure}

We carried out computations with the
Shortest Path Algorithm (see Appendix A) for $q = $2, 3, 4, 5 states,
with the field strength $h_r =  0.05$. It is easy to see that
${P_{\sigma}(q)}$ deviates some from the
${P}_{\sigma}^{nc}(q)$. Evidently, the system prefers to have the third
domain different from the first one. One can put this in two ways: 
creating a 121 configuration costs more energy, on one hand,
and on the other hand as noted before one can add a domain at the
expense of a single domain wall between two others (131 becoming 1231).
\begin{figure}
\includegraphics[width=\linewidth]
{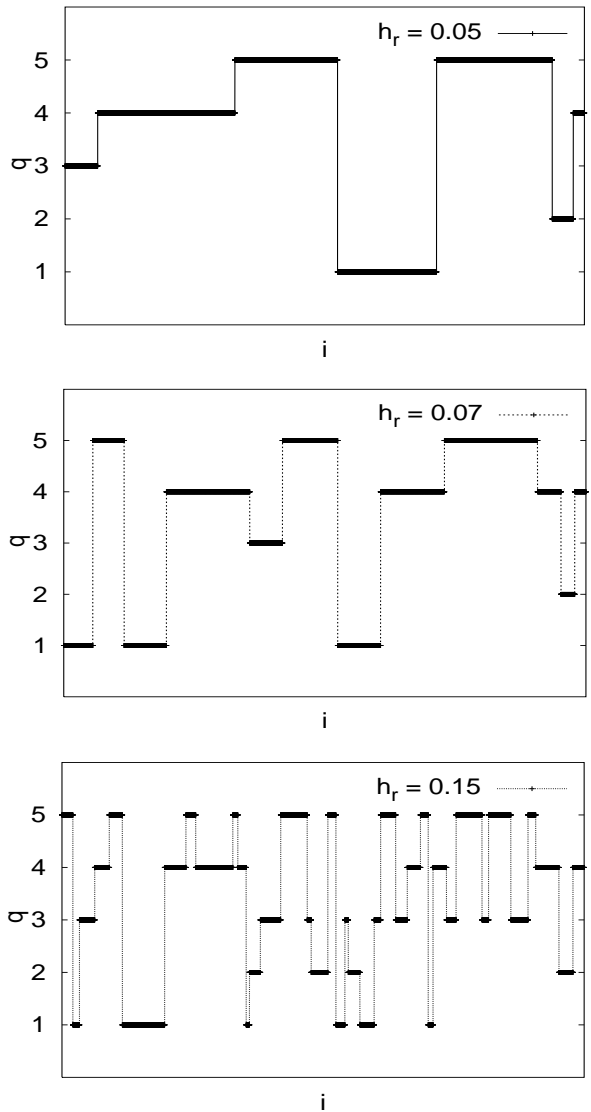}
\caption{Evolution of groundstate with increasing disorder strength $h_r$ for $q$ = 5.}
\label{fig2}
\end{figure}
  
The evolution of the domain structure with increasing field amplitude $h_r$ 
differs fundamentally from the same phenomena in the RFIC. New
domains can be created in-between two earlier ones at a smaller
cost ($2J$ instead of $4J$). Figure \ref{fig2} depicts the evolution of the
GS domains for a $q = 5$ states sample with identical configuration of
the random direction of the local fields but increasing disorder
$h_r =$ 0.05, 0.07, 0.15. The large domains are split into smaller 
fragments, mostly at their boundaries.

\subsection {Distribution of domain length and Zeeman energy}
To compare with the $q=2$ or RFIC case we now used a larger system size
of $L$ = 100000, with $n = 10^5$  samples.
Figure \ref{fig3} shows the numerical results for the accumulated Zeeman 
energy $H_Z$ as a function of domain length $l_d$ for different field 
amplitudes $h_r$. It easy to see that all $\langle H_Z \rangle$ 
versus $ l_d $ curves can be asymptotically 
(except for the smaller $l_d$ region) fitted to 
a straight line of the form\cite{Schroeder01}

\begin{equation}
\langle H_Z \rangle = qJ + d_qh_r^{\gamma}l_d.
\label{eq4}
\end{equation}
Here the coefficient $d_q$ grows roughly logarithmic for $q < 15$. 

Figure \ref{fig4}  shows the mean domain length $\langle l_d \rangle$ 
for different values of $q$ as a function of the field amplitude $h_r$. 
The data along the x-axis for $q=$ 3, 4, 5 are shifted by factors 2, 4
and 8 respectively. In the limit $h_r \ll J$ the data fits quite well 
 with the Imry-Ma scaling\cite{Imry75} { $\langle l_d
 \rangle\sim1/h_r^{2}$. The prefactor of the domain length is (see the
 inset of Figure \ref{fig4}), except for $q = 3$, linearly dependent on
 $q$. Figure \ref{fig5} finally shows the scaling plots of the
 probability distribution P($l_d$) of the domain lengths. Apart from the
 initial part, the distribution decays exponentially. We may therefore
 conclude that in spite of the slight correlations in the domain
 structure, the RFPC obeys the same scalings in the GS as the RFIC.

\begin{figure}\hspace{-3ex}
\includegraphics[width=\linewidth]
{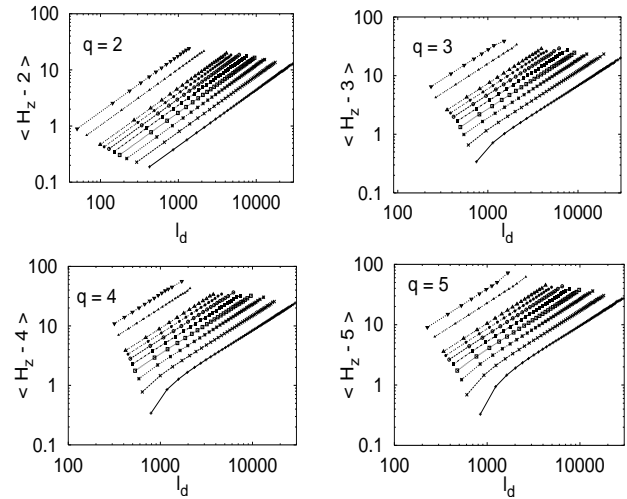}
\caption{Log-log plot of the mean Zeeman energy as a function of domain 
length $l_d$. The disorder strength $h_r =$
0.20, 0.15, 0.10, 0.09, 0.08, 0.07, 0.06, 0.05, 0.04, 0.03 
from top to bottom in all the subplots. }
\label{fig3}
\end{figure} 

\begin{figure}\hspace{-3ex}
\includegraphics[width=\linewidth]
{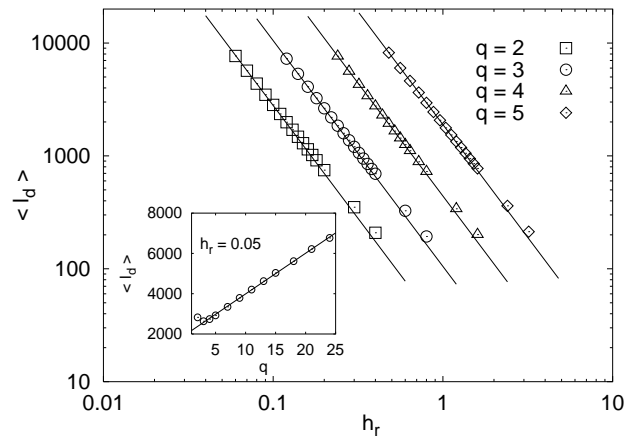}
\caption{Mean domain length as a function of $h_r$. The straight lines
 are  fits to the function $f(x) = ax^{-2}$ where the coefficient 
$a = $6.99, 6.57, 6.89 and 7.39 respectively for $q$ = 2, 3, 4 and 5. 
The fit works well for smaller $h_r$ but deviates for larger amplitudes. 
{\bf Inset:} The average domain length $\langle l_d \rangle$  as a function 
of $q$. The straight line is a linear fit with slope 201.88.}
\label{fig4}
\end{figure}

\begin{figure}\hspace{-4ex}
\includegraphics[width=\linewidth]
{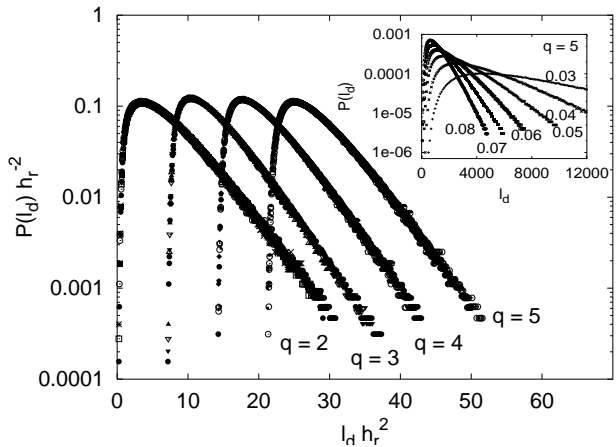}
\caption{Scaling plot of the domain length probability distribution $P(l_d)$.
The data are shifted by 7, 14, 21 units along the x-axis for $q =$ 3, 4, 5,
respectively. {\bf Inset:} 
Unscaled plot of $P(l_d)$ for $q = 5$. The labels refer to the
disorder strength $h_r$. }
\label{fig5}
\end{figure}

\subsection{Response to constant external field}
In the presence of an external field favoring one of the spin
orientations via an additional term
$h\sum_i(q\delta_{\vec{\sigma_i},\vec{\alpha_i}} - 1)$ in the
Hamiltonian($\vec{\alpha_i}$ arbitrary but fixed and the same for all
sites), the magnetization becomes of course non-zero. The
ground state magnetization is computed as
\begin{equation}
m_a = {\frac{1}{N}}\sum_{i = 1}^N\left(\frac{q\delta_{\vec{\sigma_i},
\vec{a}} - 1}{q - 1}\right).
\label{eq5}
\end{equation}
in a particular direction $\vec{a}$ (here $a\in [1,..,q])$ for $q$ 
different states. 
Now for a small field strength ($h_r \ll J$) in an infinite system one
has
\begin{equation}
{m}_{\infty}(h) \sim h.
\label{eq6}
\end{equation} 
For a finite system the corresponding finite-size scaling relation can be 
obtained as follows\cite{Rieger96,Bray85}. In the GS the Zeeman Energy of a block of spins of size
$L$ is 
\begin{equation}
E_Z \sim Lh.
\label{eq7}
\end{equation}
Further the random field (variance $h_r$) energy of the block is
\begin{equation}
E_{RF} \sim L^{1/2}h_r.
\label{eq8}
\end{equation}
For a non-zero $h$ one would expect $m_L(h) h^{-1}$ to be a function of 
the dimensionless ratio of $E_Z$ and $E_{RF}$ only, yielding
\begin{equation}
m_L(h) = h \tilde{m}(L^{1/2}h/h_r)
\label{eq9}
\end{equation}
with
\begin{eqnarray}
\tilde{m}(x) \sim 
\left\{\begin{array}{cl}
\mbox{const.} & \mbox{for $x \gg 1$}\\
\\
1/x & \mbox{for $x \ll 1$}
\end{array}\right.
\label{eq10}
\end{eqnarray}
Eq. (\ref{eq9}) can once again be written as,
 
\begin{equation}
m_L(h) = L^{-1/2}\bar{m}(L^{1/2}h/h_r)
\label{eq11}
\end{equation}
with $\bar{m}(x) = $ const. for $ x\gg1$ and $\bar{m}(x) = 1/x$ for $x \ll 1$.

\begin{figure}[t]\hspace{-3ex}
\includegraphics[width=\linewidth]
{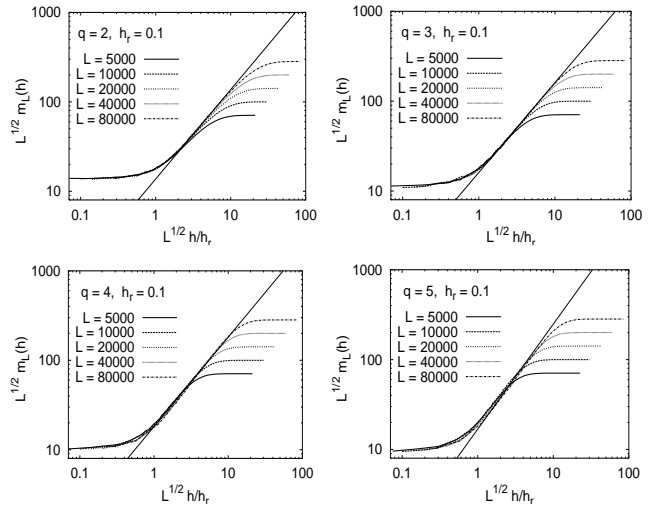}
\caption{Ground state magnetisation for $q =$ 2, 3, 4, 5, respectively. 
The exponent $\nu = 1.0$(for all $q$) follows from the fit to 
$ f(x) = bx^{\nu}$, which is also verified by the data collapse 
according to Eq. (\ref{eq10}).}
\label{fig6}
\end{figure} 

The system is subjected to a constant external field $h$ in a fixed direction 
$\vec{a} = 1$ to evaluate $m_a(S,h)$ for a sample $S$ with some given 
realization of random fields. To start, $h = 0.0001$, and the GS magnetization
is computed. Then the field is increased after averaging 1000 samples, until
saturation. Here $h_r = 0.1$, $L \in \{5000,10000,20000,40000,80000\}$ while
$h$ is in the interval [0.0001, 0.03]. For smaller systems one has to
use higher values of $h_r$, since otherwise a single domain may
percolate.

\subsection{Chaos Exponent}
A slight variation of the random fields in each sample gives rise to
``chaos''. In the case of the RFIC or RFIM in general this has already been
studied\cite{Chaos,Alava98}. There the general behavior is such that for small amplitudes
$\delta$ of the local perturbation the overlap between the new and the
old GS remains considerable, and the difference is a smooth function
of $\delta$. This is easy to see if one considers the RW factorization
of the GS: a variation of the local fields gives rise to a new RW, in
superposition with the original one. In the limit $\delta \rightarrow 0$
this has only minute effects except for barely absorbing walks and barely
non-absorbing ones that can be affected if $\delta$ is large enough.Thus 
the average overlap $q$ is linear for the RFIC in $\delta$. 
Next we look at the chaos properties in the RFPC.
\begin{figure}[t]\hspace{-5ex}
\includegraphics[width=\linewidth]
{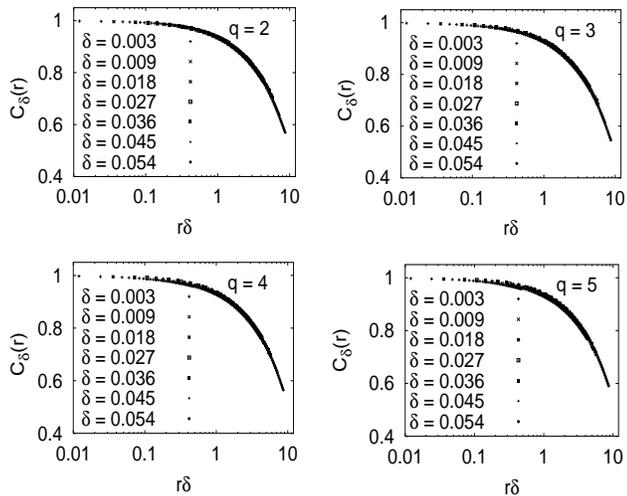}
\caption{Scaling plot of the overlap correlation function $C_\delta(r)$ versus $r\delta$. Data are averaged over 1000 samples for each perturbation strength $\delta$ with fixed system size  $L = 10^5$ and the disorder strength is fixed at $h_r = 0.05$.}
\label{fig7}
\end{figure}

To modify the initial random fields $h_i$ at each site $i$ they are
replaced by ${h_i}^{'}= h_i +\delta {\tilde{h}}$.  As before, $h_i$ and
also $\tilde{h}$ follow Gaussian distributions and the parameter
$\delta$ measures the strength of the perturbation.  Denoting
${\vec{\sigma}_{i}^{0}}$ and ${\vec{\sigma}_{i}^{\delta}}$ to be the
unperturbed and perturbed ground state at site $i$ respectively, the
ground state overlap correlation function up to a distance $r$ is defined
by
\begin{equation}
C_\delta(r) = {\left[{\frac{1}{N}}\sum_{i = 1}^{N}{\Delta({\vec{\sigma}_{i}^{0}},{\vec{\sigma}_{i+r}^{0}})}{\Delta({\vec{\sigma}_{i}^{\delta}},{\vec{\sigma}_{i+r}^{\delta}})}\right]}_{av}
\label{eq12}
\end{equation}
with

\begin{equation}
\Delta(\sigma,{\sigma}') = 2\delta_{\sigma,{\sigma}'} - 1.
\label{eq13}
\end{equation}
In the limit $N\rightarrow\infty$ one would expect a scaling form

\begin{equation}
C_\delta(r)\sim\tilde{c}(r/\xi(\delta))
\label{eq14}
\end{equation}
with

\begin{equation}
\xi(\delta) \propto \delta^{-1}.
\label{eq15}
\end{equation}

In Figure \ref{fig7} we show the results of our calculation of the GS overlap correlation function $C_\delta(r)$. $r$ is varied from 4 up to 160 for each 
perturbation strength $\delta$. The obtained data collapse agrees quite well 
with the argument above.

\section{Domains at finite temperature}
\subsection{Changes with the introduction of a finite temperature}

For $T>0$ we calculated the expectation values 
$\langle \sigma_i \rangle$  using the transfer matrix technique
\cite{Bruinsma83,Crisanti93} as described in Appendix B. 
The domains start ``melting'', i.e. the overlap with the GS,
$\langle \sigma_i \sigma_{i,GS} \rangle$,
decreases,
as the temperature is switched on from $T = 0$. 
The effect of temperature is similar to a random field perturbation
in the sense that both should play a role in particular where the
fluctuations in the random landscape (absorbing/almost absorbing walks)
make it the easiest. On the other hand the effect of temperature is not 
limited to those regions, only. Next we compare the finite temperature
and GS configurations to illuminate the $T>0$ physics.

\begin{figure}\hspace{-1ex}
\includegraphics[width=\linewidth]
{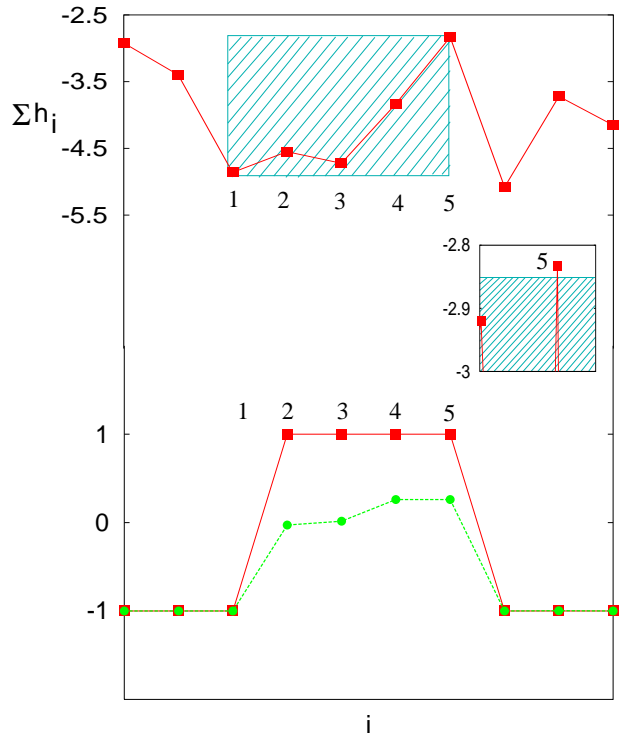}
\caption{Destruction of a GS domain for $T >0$ in the RFIC. The upper
 part of the figure shows the field sum $\sum_i h_i$ in the 
 neighbourhood of spin $i=1,...,5$, the lower part the corresponding
 ground state and a $T>0$ state. The reason why this domain melts at
 relatively low temperature is indicated in the middle part: The field
 sum of spin 5 is just outside the shaded area(see text).}
\label{fig8}
\end{figure}

In Figure \ref{fig8} we represent the destruction of a domain for a system
 of size of $L $= 400 and $h_r$ = 0.7, as the temperature is raised to 0.2 
from $T = 0$. The random field distribution is uniform in the interval 
$[-1,1]$. The zero temperature domain from $i$ = 2 to $i$ = 5 is due to a 
single absorbing excursion, depicted by the large shaded region in the 
upper part of the Figure \ref{fig8}. The inset in the middle part 
shows the sensitive region at $i$ = 5, where the field sum just crosses 
the absorbing boundary. As the temperature is increased from zero, the 
entropy gives weight to states where the whole region or much of it
is flipped, to the $\sigma = -1$ state instead of +1.
\begin{figure}\hspace{-3ex}
\includegraphics[width=\linewidth]
{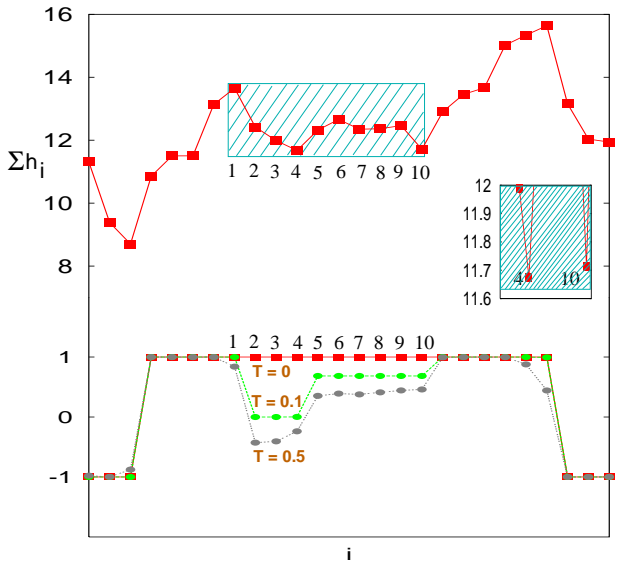}
\caption{Creation of a fluctuation for $T >0$ (RFIC). The upper part of the
 figure shows the field sum $\sum_i h_i$ in the neighbourhood of spin
 $i=1,...,10$, the lower part the corresponding ground state and two 
 $T>0$ states. The reason why this domain melts at relatively low 
 temperature is indicated in the middle part: The field sum of spin 4 
 and 10 are very close to the absorbing boundary (see text). } 
\label{fig9}
\end{figure}

Figure \ref{fig9} shows the melting of a GS domain at $T = 0.1$ for a
system of $L$ = 400 and strength of disorder $h_r$ = 1.0 (the
distribution of local fields is again uniform in [-1,1] ). Now the
sum-of-fields random walk in the GS from $i$ = 1 to 10 is actually part
of a non-absorbing excursion inside the embedding domain with $\sigma =
-1$. It is however susceptible to the introduction of temperature (note
the spins $i = $ 4 and 10 where the RW is closest to creating a domain,
in the inset).

At a finite temperature the sites 2, 3, 4 form a ``new domain''
 with $\sigma = -1$. This new domain contains only one part of the
RF fluctuation, from 1 to 4. 

To further investigate the melting of a GS domains at finite
temperatures, an overview of the GS and equilibrium configurations at
finite temperatures is shown in Figure \ref{fig10}. The calculations for
each temperature including the GS are carried out for the same local
field configuration. The thick dotted line corresponds to the GS whereas
the continuous lines represents the melted configurations with the
gradual increase in temperature.
 
\begin{figure}\hspace{-3ex}
\includegraphics[width=\linewidth]
{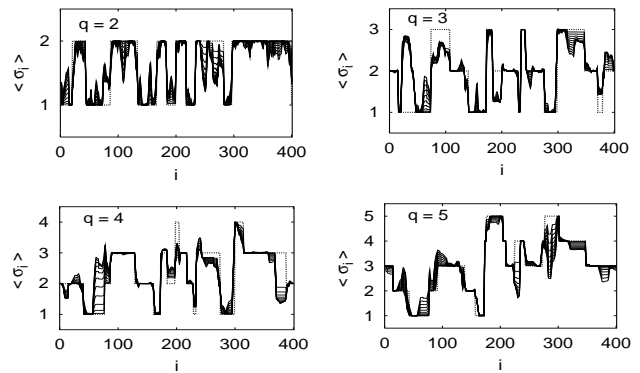}
\caption{GS of the $q$-States Random field Potts model and equilibrium states 
for $T>0$. The GS domains are represented by a thick dotted line. The range of 
temperature goes up to $T = 0.7$. Between $T = 0.25$ and $T = 0.7$ 
$\Delta T = 0.05$. The system size is restricted to $L = 400$ and $h_r = 0.5$}
\label{fig10}
\end{figure}

In general, for very low temperatures there are only a few segments of spins, 
which differ from their respective GS orientations \cite{Schroeder01}, growing
with increasing temperature in height and width. At high temperatures
$(T \gg J$ and $T \gg h_r)$ the expectation value $\langle \sigma \rangle$ 
of each spin finally fluctuates around the mean value $1/q$.

\subsection{Domains at $T>0$}
To construct a domain structure at any finite temperature we focus to
the melting of individual $q$ states. It is easy to calculate the
contribution for each of $q$ states to the resultant melting by taking a
single state in the diagonal matrix, in the transfer matrix calculation,
and keeping the others zero. The maximum of these contributions will
give rise to a definite local spin state ($\sigma_i = 1, ....q)$.
Figure \ref{fig11}(a) represents the GS (straight bold line) and melted
$\langle \sigma_i \rangle$ at finite temperatures for $q = 3$. Figures
\ref{fig11}(b), (c), (d) shows the components (for each $q$) of
probability $P(\sigma_i = 1,2,3)$ at $T \geq 0$ of individual spins
$\langle \sigma_i \rangle$ at sites $i$. Consider the 1st segment of
domain in (a). It is evident that the melted domain approaches the state
$q = 3$ as confirmed by the behavior of the same segment in Figures (b),
(c) and (d).  To analyze the melting process quantitatively we
investigate the probability distributions of lengths of the melted
segments and the melting rates $\frac{\Delta m_i}{\Delta T}$, where 
$m_i= \langle \sigma_i \rangle$ for each of $q = $2, 3, 4, 5.

\begin{figure}\hspace{-1ex}
\includegraphics[width=\linewidth] {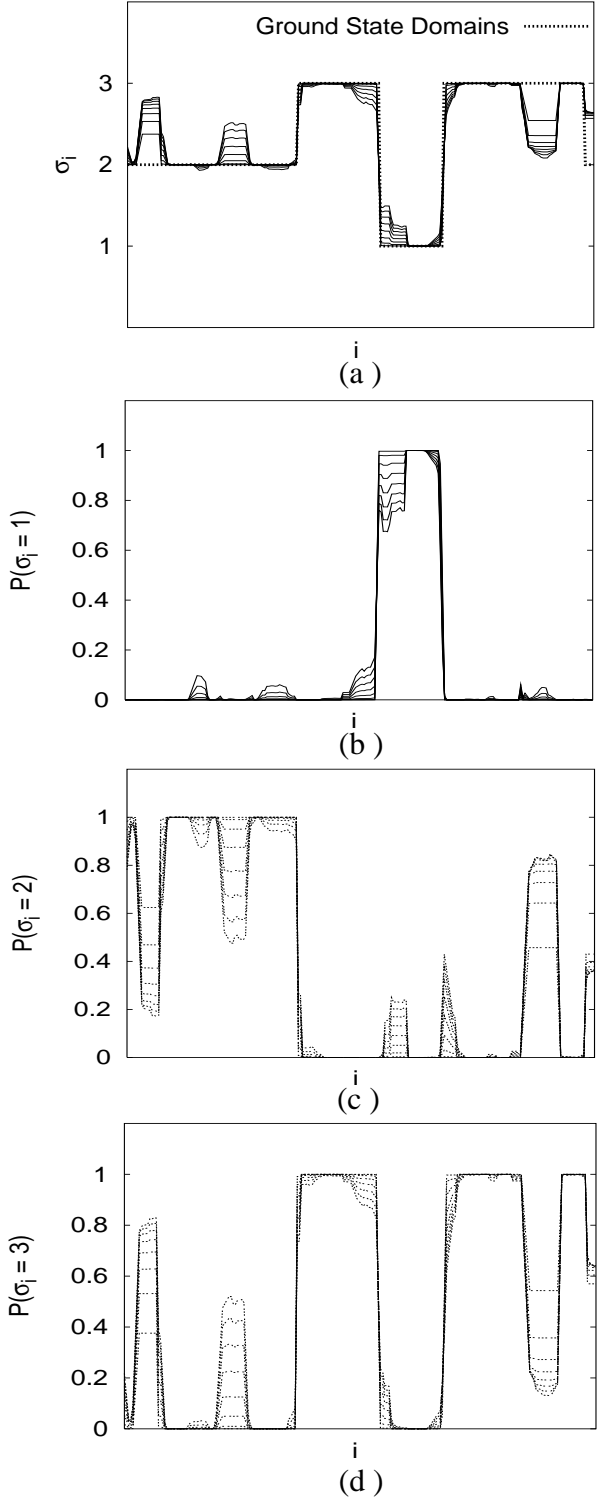}
\caption{The GS of RFPC with $q = 3$ and equilibrium states
($\langle\sigma_i\rangle$ is the thermal expectation value of the spin
value at site $i$) for $T>0$.  Within the interval $0.1\leq T \leq
0.45$ the temperature is raised by $\Delta T = 0.05$. (b), (c), and (d)
shows components of melting probabilities for $q = 1 \dots 3$.}
\label{fig11}
\end{figure} 

We also compute the finite temperature distribution of domain lengths. 
The simulations are for different temperatures with a system of fixed 
$L = 300$ and disorder strength $h_r = 0.7$. The distribution of local random 
fields is uniform in [-1,1]. The disorder, temperature,
 and system size are restricted by the fact that one has to compute products 
of $L(q\times q)$ transfer matrices. One also prefers to avoid Gaussian 
disorder instead of one with a bounded support. It is clear from 
Figure \ref{fig12}, that except for the initial part, the probability 
distribution  $P(l_T)$ varies exponentially with $l_T$, similarly to GS 
domains. With an increasing temperature the decay in the tail becomes
faster, indicating that correlations diminish as expected.

 \begin{figure}\hspace{-3ex}
\includegraphics[width=\linewidth]
{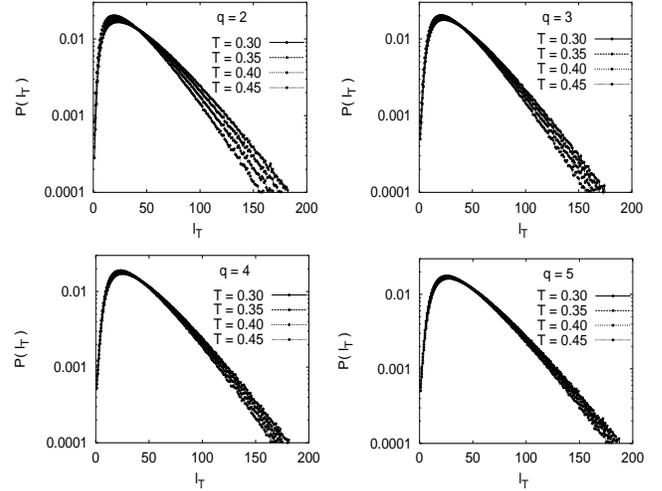}
\caption{Probability distribution of finite $T$ domains $P(l_T)$ for 
$q$ = 2, 3, 4, 5. The data are averaged over $1065$ samples.}
\label{fig12}
\end{figure} 

We define the probability distribution of melting rates 
$\frac{\Delta m_i}{\Delta T}$ as,
\begin{equation}
P(\frac{\Delta m_i}{\Delta T}) =\frac{N(\frac{\Delta m_i}{\Delta T})}{{n L}},
\label{eq16}
\end{equation}
with 
\begin{equation}
\Delta m = |\sigma(T + \Delta T) - \sigma(T)|,
\label{eq17}
\end{equation}
where $n$ is the number of samples. We measure the change in magnetisation 
$m_i$ for each spin $\sigma_i$ when the temperature changes by $\Delta T$. 
For counting we use the resolution $\delta = 0.001$, for which 
$N(\frac{\Delta m}{\Delta T})$ is actually the number of spins $\sigma_i$ 
with $\frac{\Delta m_i}{\Delta T} \in [\frac{\Delta m}{\Delta T} -  
\frac{\delta}{2}  ,\frac{\Delta m}{\Delta T} + \frac{\delta}{2}]$.

\begin{figure}\hspace{-3ex}
\includegraphics[width=\linewidth]
{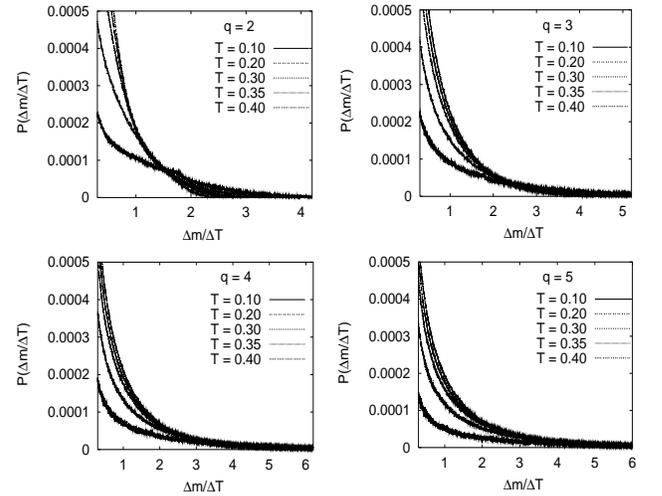}
\caption{Probability distribution $P(\frac{\Delta m}{\Delta T})$ at constant 
field $h_r = 0.7$.}
\label{fig13}
\end{figure}

Figure \ref{fig13} shows the distribution $P(\frac{\Delta m}{\Delta T})$ 
at different temperatures for $q = $2, 3, 4, 5; $\Delta T = 0.05$ is the same
for all temperatures and $10^5$ configurations. The data points for 
$\frac{\Delta m}{\Delta T}$ are not shown as in this regime we observe a 
strong maximum at $\frac{\Delta m}{\Delta T} \rightarrow 0$. For $q = 2$ there
is a small but distinct peak in the curve for $T = 0.2$ which gradually 
diminishes for higher values of $q$.

\subsection{Finite temperature lengthscales}
Finally we investigate how the local regions of the magnetization
evolve at finite $T$\cite{Schroeder01}. 
We calculate the average length $l_m(l_m = \langle l_T 
\rangle)$ from the finite temperature configurations. Figure \ref{fig14} 
represents how this lengthscale changes with temperature, if one
first scales away the $T = 0$ dependence on the field. 
Note that the temperature is restricted so that $T \ge 0.09$. 
A further collapse with the right combination of $h$ and $T$ makes it 
possible to observe an universal scaling function for $l_m$ so that
\begin{equation}
l_m = l_d F(\frac{h^{\nu}}{T}),
\label{eq18}
\end{equation}
where the scaling function $F \rightarrow 1$ as $T \rightarrow 0$. 
The exponent
\begin{equation}
\nu \cong 2/3
\label{eq19}
\end{equation}
does not change with $q$, and thus one has the same scaling
for $q>2$ as for the RFIC. The implication is basically the
same: one has per each domain a number of ``easy'' excitations
that allow the magnetization to change a lot from the GS. The
presence of such almost degenerate regions is not affect by
the exact value of $q$.

\begin{figure}\hspace{-3ex}
\includegraphics[width=\linewidth]
{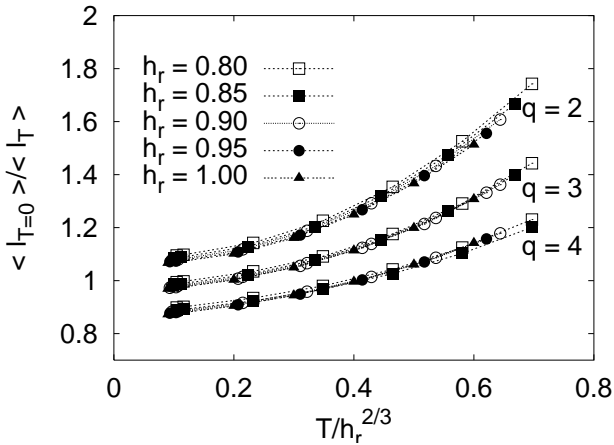}
\caption{Scaling plot of the average $T>0$ domain length for different 
$h_r$. The data are shifted along the y-axis by -0.1 and -0.2 units for 
$q = $3, 4 respectively. $L = 300$ and the data are averaged over 50000 
configurations.}
\label{fig14}
\end{figure}

\section{RFIC with correlated random fields}
We have already seen that for an uncorrelated random field distribution
the average domain length asymptotically follows the Imry-Ma argument,
i.e. $\langle l_d \rangle \sim 1/h_r^{-2}$. This should change for
spatially correlated random fields. Now we obtain an exponent $\gamma$ 
with $\langle l_d \rangle \sim 1/h_r^{-\gamma}$ where $\gamma$ approaches 2 
as the spatial correlations decrease. Consider a power-law correlation 
function of the form
\begin{equation}
\langle h_i h_{i+r}\rangle \sim r^{-\rho}.
\label{eq20}
\end{equation}
To use an analogy of the Imry-Ma argument, the random field energy of a 
domain of length $l_d$ is expressed as
\begin{equation}
E_{RF} \propto \sqrt{\sum_{i,j = 1}^{l_d}h_i h_j}.
\label{eq21}
\end{equation}
Making use of the correlation function (21) we readily obtain,
\begin{equation}
E_{RF} \simeq \sqrt{h_r^2\left(\frac{l_d^{2-\rho}}{1-\rho}\right)} 
\label{eq22}
\end{equation}
with  $0<\rho<1$. The domain wall energy is as usual 
\begin{equation} 
E_{DW} \simeq J
\label{eq23}
\end{equation}
To create domains, in the GS, $E_{RF}$ and $E_{DW}$ are of the same order 
of magnitude, thus
\begin{equation}
l_d \simeq h_r^{-\gamma}
\label{eq24}
\end{equation}
where the exponent
\begin{equation}
\gamma = \frac{2}{2-\rho}.
\label{eq25}
\end{equation}
for $\rho<1$. For $\rho>1$ this Imry-Ma type argument predicts the
irrelevance of the correlations in the disorder, i.e. $\rho=2$.
\begin{figure}\hspace{-3ex}
\includegraphics[width=\linewidth]
{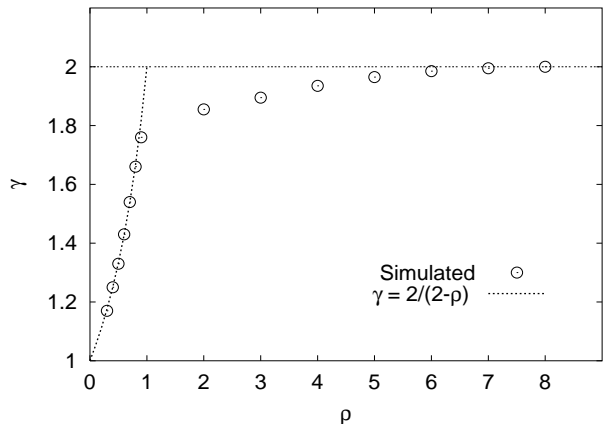}
\caption{Scaling exponent $\gamma$ of  $p(l_d) = h_r ^{ -\gamma}
P(h_r ^{ \gamma}l_d)$ against $\rho$. $\rho$ is the exponent of the 
correlated random field distribution (see text).}
\label{fig15}
\end{figure}

We computed the probability distribution $P(l_d)$ on a large system of
size $L = 65536$, with the local, correlated, Gaussian random fields
generated by the Noise Construction Algorithm\cite{Pang95}.
Statistical averaging was performed over $10^5$ samples. As in the
case of uncorrelated fields $P(l_d)$ is again exponential for all
values of $\rho$ we studied. The disorder $h_r$ was varied for each
value of $\rho$, to calculate the exponent $\gamma$ from the data
collapse of $P(l_d)$ versus $l_d$. $\gamma$ deviates significantly
from the predicted value as $\rho$ approaches unity, but finally
approach the value 2 for $\rho\rightarrow \infty$. The deviation might
be due to logarithmic corrections to the Imry-Ma prediction
(\ref{eq25}) close to the critical value $\rho = 2$. The same
properties can also be verified for the RFPC, if the random field
magnitudes and their directions obey the same correlations.

\section{Conclusions}

It is a fortunate fact that one can solve numerically, but exactly,
for the groundstates of the RFPC. For the RFIM one
has access to powerful graph optimization methods that work in $d$
arbitrary, but for $q>2$ it can actually be shown that the problem
is NP-complete for $d>1$ \cite{Preissmann97}. Here we have used this to study the Potts
chains, augmented with random walk considerations and with transfer
matrix calculations.

As it is already known that the RFIC groundstate is, essentially,
separable to independent regions and what the consequences are
it is most natural to compare the $q>2$ behavior to the RFIC 
physics. It turns out that essentially all more general scaling
behaviors, whether pertaining to the GS or to the finite temperature
states, follow similar laws. This is true, in groundstates,
for the Zeeman energy, for the form of the domain distribution,
and for the average domain length. At $T>0$ we have demonstrated
that the change from the GS configurations follows a similar course
for $q$ arbitrary. Finally also for GS chaos the value of $q$ has
no essential importance. 

There are of course slight differences since the local domain
structure has correlations due to the ``decorations'' of large
domains with smaller ones that utilize the fact that only one
domain wall needs to be created if e.g. induced by increasing
the disorder strength. These are of secondary importance
for the scaling properties. 

The greatest deviations from the expected behavior we observed
in the case of the RFIC with correlated fields, where the scaling
exponent of the domain length with disorder strength does not seem
to take the expected value $\gamma=2$ (but is slightly less) for
weakly correlated fields. This may be due to corrections to scaling,
or to the detailed properties of a RW picture, similarly to the
uncorrelated case, when the walks are fractional Brownian motions.

In the case of the RFPC it is possible to propose a number of
topics to study. Eg. the generalizations of the multifractal
magnetization distributions of the RFIC seem mathematically
interesting. Also the detailed properties of the ``melting''
of the GS might merit further study.

\section*{ACKNOWLEDGEMENTS}
We are grateful to G. Schr\"oder for contributing to the early stages
of this work, in particular for the design of the graph on which the
algorithm for finding the ground states of the RFPC operates. This
work has been partially supported by the Deutsche
Forschungsgemeinschaft(DFG) and the European Science Foundation (ESF)
Sphinx program.
MJA acknowledges the contribution of the Center of Excellence program
of the Academy of Finland.

\begin{appendix}
\section{}
\begin{figure}[t]\hspace{-3ex}
\includegraphics[width=\linewidth]
{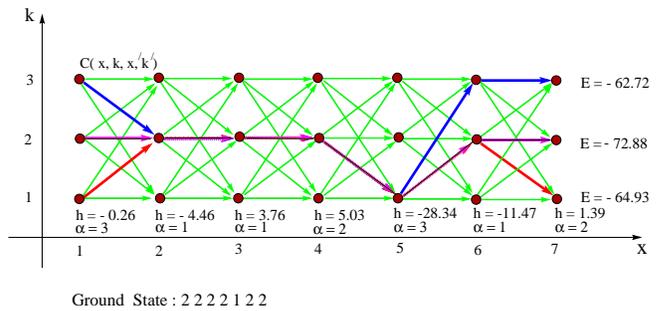}
\caption{The graph corresponding to the Shortest Path Method for the RFPC problem.}
\label{fig16}
\end{figure}
The ground state of $q$-State RFPC can be calculated exactly by mapping 
onto a shortest path problem on an acyclic graph.
The graph $G(L,q)$ corresponding to a $q$-State RF Potts chain of length $L$ 
consists of ($L$+1)$\times q$ nodes which we consider to be arranged like
presented in figure 1 and denoted by ($x,k$) with $x = 1,....,L$ and 
$k = 1,....,q$. Each node ($x,k$) is connected to all of its right 
neighbours, i.e. to all nodes ($x+1,k'$) with $k' = 1,....,q$ by a 
directed one-way edge with cost
\begin{equation}
c_{(x,x+1,{k},{k}')} = -J(q\delta_{\vec{k},\vec{k'}} - 1) - h_{x+1}(q
\delta_{\vec{k},\vec{\alpha}} - 1).
\label{eq26}
\end{equation}
Fixing the spin at $x =$ 0 to be in state $k$ the groundstate of RFPC then 
corresponds to a shortest path from ($1,k$) to ($L,k$) in $G$($L,k$). To 
apply periodic boundary conditions it is necessary to solve $q$ shortest path 
problems, one for each of the $q$ (with $(q,q)$ spins at sites $x =$ 1 and 
$x = L$+1) possible states. One of these shortest paths with minimum energy 
gives rise to the optimal groundstate.

Since the underlying graph (Fig.\ \ref{fig16}) is acyclic no cycle, in
particular none with negative weight, can occur and Dijkstra's algorithm
for finding the shortest path in a weighted graph can be applied
although not all edge weights are positive. The running time of this
algorithm is linear both in number $n$ of the nodes and in the number
$m$ of the edges, thus the overall complexity to find the groundstate of
the RFPC with periodic boundary condition is $\mathcal{O}$($q^2L$).

\section{}
For temperature $T > 0$ we generalize the {\em Transfer Matrix Method}
\cite{Crisanti93} for the $q$-States RFPC. The calculation of the
partition function ${\mathcal{Z}}_N$ can be reduced to the problem of finding 
the product of $N$ random matrices:
\begin{equation}
{\mathcal{Z}}_N = \sum_{\{\sigma\}}exp(-\beta{\mathcal{ H}}{\sigma}) = 
Tr\left(\prod_{i=1}^{N}{{\bf L}^{(i)}}\right)
\label{eq27}
\end{equation} 
where
\begin{equation}
{\bf L}^{(i)}_{j,k} = \left[\begin{array}{c}
\exp[\beta J(q\delta_{\vec{\sigma_j},{\vec{\sigma_k}}}-1)+\beta h_i(q
\delta_{\vec{\sigma_j},{\vec{\alpha_i}}}-1)]
\end{array}\right]_{q\times q},
\label{eq28}
\end{equation}
Here ($j,k=1,...,q$).

Using this expression of partition function the expectation value $\langle
 \sigma_r \rangle$ for each spin can be expressed as,
\begin{equation}
\langle \sigma_r \rangle = \frac{\sum_{\{\sigma\}}\sigma_r exp{(-\beta{
\mathcal{H} })}}{{\mathcal{Z}}_N}
\label{eq29}
\end{equation}

It is easy to see from Eq. (\ref{eq17}) that 
\begin{equation}
\langle \sigma_r \rangle = \frac{{Tr\left[\left(\prod_{i=1}^{N}
{\bf L}^{(i+r-1)}\right){\bf S}\right]}}{Tr\left(\prod_{i=1}^{N}
{\bf L}^{(i+r-1)}\right)}
\label{eq30}
\end{equation}
with  ${\bf {S}}_{ij} = i.\delta_{ij}$.

The only computational effort consists in calculating the product of $N$ 
$q\times q$ transfer matrices. Since the elements of ${\bf L}^{(i)}$ become
very small for low temperatures, arbitrarily small temperatures can not be 
considered. However, the admissible temperature interval is sufficient for 
our investigations.
\end{appendix}

\end{document}